\documentclass[letterpaper]{jpconf}
\usepackage{graphicx}
\usepackage{amsmath}
\begin{document}
\title{Competing magnetic interactions in CeNi$_{9-x}$Co$_x$Ge$_4$}

\author{L Peyker$^1$, C Gold$^1$, W Scherer$^1$, H Michor$^2$ and E-W Scheidt$^1$}

\address{$^1$ CPM, Institut f\"{u}r Physik, Universit\"{a}t Augsburg, 86159
Augsburg, Germany}
\address{$^2$ Institut f\"{u}r Festk\"{o}rperphysik, Technische Universit\"{a}t Wien,
1040 Wien, Austria}

\ead{ludwig.peyker@physik.uni-augsburg.de}

\begin{abstract}
CeNi$_9$Ge$_4$ exhibits outstanding heavy fermion features with
remarkable non-Fermi-liquid behavior which is mainly driven by
single-ion effects. The substitution of Ni by Cu causes a reduction
of both, the RKKY coupling and Kondo interaction, coming along with
a dramatic change of the crystal field (CF) splitting. Thereby a
quasi-quartet ground state observed in CeNi$_9$Ge$_4$ reduces to a
two-fold degenerate one in CeNi$_8$CuGe$_4$. This leads to a
modification of the effective spin degeneracy of the Kondo lattice
ground state and to the appearance of antiferromagnetic (AFM) order.
To obtain a better understanding of consequences resulting from a
reduction of the effective spin degeneracy, we stepwise replaced Ni
by Co. Thereby an increase of the Kondo and RKKY interactions
through the reduction of the effective $d$-electron count is
expected. Accordingly, a paramagnetic Fermi liquid ground state
should arise. Our experimental studies, however, reveal AFM order
already for small Co concentrations, which becomes even more
pronounced with increasing Co content $x$. Thereby the modification
of the effective spin degeneracy seems to play a crucial role in
this system.
\end{abstract}

\section{Introduction}
Due to the wide variety of their ground states, Ce based ternary
intermetallic compounds have attracted large attention during the
last years. Inherent to these materials is the existence of
localized $4f$ electrons (Ce$^{3+}$)  at high temperatures. In the
low temperature limit, the competition between Kondo and RKKY
interactions results in different ground states depending on their
relative magnitudes~\cite{Stewart2001}: The formation of long-range
magnetic order depends quadratically on the dimensionless coupling
parameter $N(E_\mathrm{F})J_0$, while the development of a local
Kondo ground state depends exponentially on $N(E_\mathrm{F})J_0$
\cite{Doniach1977}. The dimensionless effective exchange coupling
parameter $N(E_\mathrm{F})J_0$ correlates the exchange interaction,
$J_0$, between $4f$ localized magnetic moments and conduction
electrons with the electronic density of states at the Fermi level,
$N(E_\mathrm{F})$. Accordingly, small values of $N(E_\mathrm{F})J_0$
favor long range magnetic order, while for large values of
$N(E_\mathrm{F})J_0$ a paramagnetic Kondo-screened ground state is
anticipated. At the borderline between these two regimes non
Fermi-liquid (NFL) behavior
has been observed.\\
Apart from this classical scenario other mechanisms, e.g., via a
change of the effective spin degeneracy $N$ may occur. The latter
was analyzed by Coleman \cite{Coleman1983}. This additional
mechanism seems to be relevant for the ground state of the heavy
fermion system CeNi$_9$Ge$_4$ \cite{Peyker2009}. In the substitution
series CeNi$_{9-x}$Cu$_x$Ge$_4$ a reduction from a quasi-quartet
ground state in CeNi$_9$Ge$_4$ to a two-fold degenerated one in
CeNi$_8$CuGe$_4$ leads to antiferromagnetic (AFM)
order\cite{Peyker2009}. This is also benefitted by a reduction of
the effective exchange coupling
$N(E_\mathrm{F})J_0$ due to an enhanced effective $d$-electron count.\\
In the present work, the hybridization strength in CeNi$_9$Ge$_4$ is
tuned contrary through non-isoelectronic Ni/Co substitution to study
the influences of the effective spin degeneracy in more detail.\\

\section{Sample preparation and characterization}
For the synthesis of the samples of CeNi$_{9-x}$Co$_x$Ge$_4$
materials of high purity were used: Ce: 4N; La: 3N8 (Ames MPC); Ni:
4N5; Co: 4N8; Ge: 5N. They were prepared by arc melting under argon
atmosphere followed by an annealing process at $950^\circ$\,C for
two weeks inside an evacuated quartz tube. Less than $0.5$\,\%
weight loss occurred during the melting process. Optical emission
spectroscopy in an inductively coupled plasma (ICP-OES) and energy
dispersive X-ray spectroscopy (EDX) analysis indicated that the
samples used in this work are essentially single phase. X-ray powder
diffraction experiments revealed that all samples crystallize in the
LaFe$_9$Si$_4$-type structure (tetragonal spacegroup $I4/mcm$). For
the whole concentration range, the replacement of Ni by Co does not
lead to any significant changes in the lattice parameters and
therefore also not in the unit cell volume. Thus, volume effects in
CeNi$_{9-x}$Co$_x$Ge$_4$ due to the substitution,
i.e. chemical pressure, are hardly relevant.\\

\section{Experimental results}
\begin{figure}
\begin{center}
\includegraphics[width=30pc]{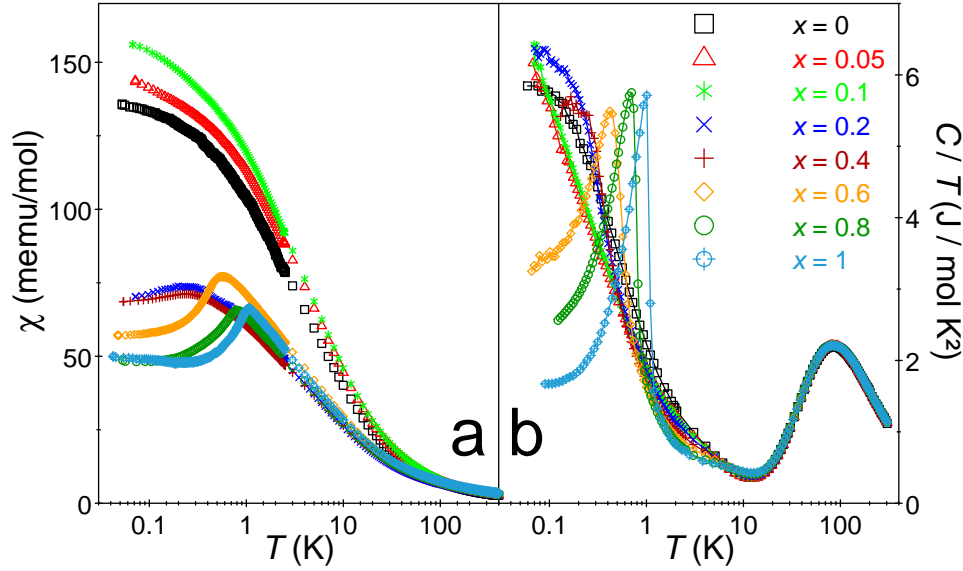}
\end{center}
\caption{\label{fig1}(Color online) (a) The magnetic susceptibility
$\chi$ and (b) the specific heat divided by temperature $C/T$ of
CeNi$_{9-x}$Co$_x$Ge$_4$ in semi-logarithmic plots. AFM transitions
are evident for $x\geq0.2$.}
\end{figure}

Figure\,\ref{fig1}\,a presents the temperature dependent
magnetic susceptibility $\chi$ of CeNi$_{9-x}$Co$_x$Ge$_4$.
The magnetic behavior of the parent compound CeNi$_9$Ge$_4$ is
discussed in more detail in \cite{Michor2006}. The Co substituted
samples follow a simple modified Curie-Weiss-type law,
$\chi(T)=C/(T-\Theta)+\chi_0$, above 100\,K . From a least-square
fit the following parameters were obtained: i) the paramagnetic
Curie-Weiss temperature $\Theta$ around $-40$\,K, ii) the
temperature independent susceptibility contribution $\chi_0$ between
0.9\,memu\,mol$^{-1}$ for $x=0.05$ and 1.6\,memu\,mol$^{-1}$ for
$x=1$, iii) the Curie constant $C$ corresponding to an effective
paramagnetic moment between $2.5$\,$\mu_\mathrm{B}$ and
$2.6$\,$\mu_\mathrm{B}$, which is in line with the theoretical value
of $2.54$\,$\mu_\mathrm{B}$ for a Ce$^{3+}$ ion. An itinerant
paramagnetic Co sublattice as observed in La- and CeCo$_9$Ge$_4$
\cite{Gold2010} is not observed. For smallest Co concentrations
$x\leq0.1$ the substitution of Ni by Co causes a moderate increase
of the low temperature susceptibility as compared to the parent
compound CeNi$_{9}$Ge$_4$. This may be caused by the influence of
substitutional disorder leading to partial reduction of the AFM
intersite coupling or to some ferromagnetic correlations. Between $x
= 0.1$ and $x =0.2$ the low temperature susceptibility decreases
dramatically. This huge change in the magnetic properties can not
only be caused by a marked increase of the Kondo screening but also
by a reduction of the effective spin degeneracy. On further
increasing $x$ the development of sharp cusps in $\chi(T)$ finally
indicates phase transitions towards AFM order. CeNi$_{8}$CoGe$_4$
exhibits magnetic ordering below $T_\mathrm{N}\approx1$\,K.\\
These observations are corroborated by specific heat results.
Figure\,\ref{fig1}\,b shows the specific heat divided by temperature
$C/T$ of CeNi$_{9-x}$Co$_x$Ge$_4$ between 60\,mK and 300\,K on a
semi-log scale. Already a small amount of Co ($x\leq0.1$) causes a
$C/T \propto -\ln(T)$ divergence of the
\begin{figure}
\begin{center}
\includegraphics[width=30pc]{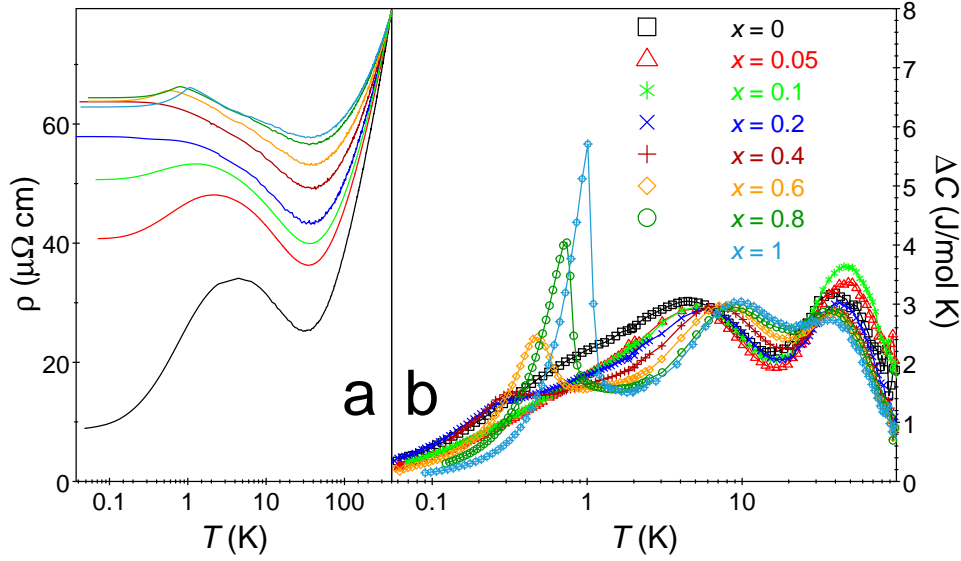}
\end{center}
\caption{\label{fig2}(Color online) (a) The electrical resistivity
$\rho(T)$ of CeNi$_{9-x}$Co$_x$Ge$_4$ normalized at 300\,K to that
of LaNi$_9$Ge$_4$ in a semi-logarithmic plot. (b) Temperature
dependence of the magnetic specific heat $\Delta C$ of
CeNi$_{9-x}$Co$_x$Ge$_4$ in semi-logarithmic representation.}
\end{figure}
Sommerfeld coefficient which holds over more than one decade in
temperature down to $T=70$\,mK. This reveals a temperature dependent
Sommerfeld-Wilson ratio, $R\propto\chi_0/\gamma$, which has been
discussed in terms of crystal field effects and a fourfold
degenerated ground state in the case of CeNi$_9$Ge$_4$
\cite{Scheidt2006, Anders2006}. In addition, the divergence of the
Sommerfeld coefficient indicates the vicinity to a QCP transition
which may occur around $x = 0.15$. At higher Co concentrations
susceptibility and specific heat reveal, initially a flattening
($x=0.2;0.4)$ and then a peak $x\geq0.6$ indicating an AFM
transition. Additionally, the low temperature limit of $C/T$
decreases with higher $x$ values, indicating an enhancement of the
Kondo coupling.\\
In Figure\,\ref{fig2}\,a the electrical resistivity of
CeNi$_{9-x}$Co$_x$Ge$_4$ is displayed. The classical Kondo lattice
behavior of CeNi$_{9}$Ge$_4$ which is observed, is explicit
illustrated in \cite{Peyker2009}. The two samples with a small Co
concentration ($x=0.05$ and $x=0.1$) exhibit a similar behavior but
with enhanced residual resistances at low temperature. Additionally,
with increasing Co content, substitutional disorder shifts the Kondo
coherence maximum to lower temperatures. On further increase of $x$
the Kondo lattice behavior disappears and the residual resistance
increases further. The cusps in the electrical resistivity for
$x\geq0.6$ indicate the AFM transition, already observed in magnetic
susceptibility and specific heat data.\\
In order to study the crossover from single-ion Kondo behavior
towards long range AFM order, the magnetic contribution to the
specific heat $\Delta C$ (Fig\,\ref{fig2}\,b) was extracted by
subtracting the total specific heat of the corresponding La
compounds with unoccupied $4f$ states. For CeNi$_9$Ge$_4$ two
pronounced maxima around 5 und 35\,K are obtained. The first
Schottky-like anomaly is associated with a quasi-quartet ground
state characterized by a subtle splitting into two doublets
$\varGamma_7^{(1)}$ and $\varGamma_7^{(2)}$ which interacts with the
Kondo screening on the same energy scale \cite{Michor2006}. The
second maximum is associated with the $\varGamma_6$ CF doublet. With
increasing Co concentration the upper maximum remains almost
constant, while the lower maximum shifts to higher temperatures
(10\,K for $x=1$). This refers to a reduction of the effective spin
degeneracy $N$ like it was found in CeNi$_{9-x}$Cu$_x$Ge$_4$
\cite{Peyker2009} and appears to be the driving force for the AFM
ordering. For $x \geq 0.2$ the
AFM phase transition at low temperatures is observed.\\

\section{Conclusions}
Studying the ground state evolution in CeNi$_{9-x}$Co$_x$Ge$_4$, we
found already for small Co concentrations $x\leq0.1$ a logarithmic
divergence of the Sommerfeld coefficient $\gamma\simeq C/T$, as it
is typical for a QCP transition. In contrast, the magnetic
susceptibility of these compounds shows a flattening instead of the
logarithmic divergence which would be expected in case of a
temperature independent Sommerfeld-Wilson ratio $\chi_0/\gamma$. A
further increase of the Co content $x\geq0.2$ leads to a significant
pronounced increase of the Kondo temperature but nonetheless AFM
order appears in this system. The latter is counter-intuitive to the
usual Doniach picture and seems to result from a reduction of the
effective spin degeneracy $N$. The fact that a similar scenario is
observed also in CeNi$_{9-x}$Cu$_x$Ge$_4$ \cite{Peyker2009}, where
the $3d$ electron count changes in the opposite way, clearly
suggests that the reduction of the effective spin degeneracy away
from the paramagnetic quasi-quartet ground state in CeNi$_9$Ge$_4$
plays a crucial role in the formation of long range
AFM order in the system CeNi$_{9-x}TM_x$Ge$_4$ ($TM =$ Cu, Co).\\

\section{Acknowledgments}
This work was supported by the Deutsche Forschungsgemeinschaft (DFG)
under Contract No. SCHE487/7-1/2.

\providecommand{\newblock}{}

\end{document}